\documentclass[10pt,letterpaper]{article}
\usepackage{opex3}
\usepackage{cite} 
\usepackage[utf8]{inputenc} 
\usepackage{amssymb}
\usepackage{amsmath}
\usepackage{subfigure}
\usepackage{amsmath}
\usepackage{graphicx} 
\usepackage{bm}       
\usepackage{todonotes}
\usepackage[squaren]{SIunits} 
\DeclareGraphicsExtensions{.pdf,.png}

\renewcommand{\vec}[1]{\mathbf{{#1}}}
\newcommand{\pvec}[1]{\mathbf{{#1}}_\parallel }
\newcommand{\imu}{\mathrm{i}}
\newcommand{\dint}{\mathrm{d}}

\renewcommand{\Re}{\mathrm{Re}}
\renewcommand{\Im}{\mathrm{Im}}

\begin{document}

\title{Satellite peaks in the scattering of light from the two-dimensional randomly rough surface of a dielectric film on a planar metal surface}
\author{T. Nordam,$^{1}$ P.A. Letnes,$^1$ I. Simonsen,$^{1*}$ and A.A. Maradudin$^2$}
\address{
    $^1$Department of Physics, Norwegian University of Science and Technology (NTNU) \\
    NO-7491 Trondheim, Norway\\
    $^2$Department of Physics and Astronomy,
    University of California, \\
    Irvine CA 92697, U.S.A.
}
\email{$^*$Ingve.Simonsen@phys.ntnu.no}



\begin{abstract}
  A nonperturbative, purely numerical, solution of the reduced Rayleigh equation for the scattering of p- and s-polarized light from a  dielectric film with a two-dimensional randomly rough surface deposited on a planar metallic substrate, has been carried out.  It is found that satellite peaks are present in the angular dependence of the elements of the mean differential reflection coefficient in addition to an enhanced backscattering peak.  This result resolves a conflict between the results of earlier approximate theoretical studies of scattering from this system.
\end{abstract}
\ocis{
    (290.1483)   BSDF, BRDF, and BTDF;
    (290.4210)   Multiple scattering;
    (290.5825)   Scattering theory;
    (290.5880)   Scattering, rough surfaces.
}




In the earliest analytic~\cite{1} and computer simulation~\cite{2,3}
studies of the multiple scattering of light from clean one-dimensional
randomly rough surfaces of perfect conductors or of penetrable media,
the focus was on the phenomenon of enhanced backscattering.  This is
the presence of a well-defined peak in the retroreflection direction
in the angular dependence of the intensity of the light that has been
scattered incoherently (diffusely).

In subsequent work on the multiple scattering of light from
free-standing or supported films with a one-dimensional randomly rough
surface that support two or more guided waves, new effects were
discovered~\cite{springerlink:10.1140/epjst/e2010-01221-4}.
These include enhanced transmission, which is the
presence of a well-defined peak in the anti-specular direction in the
angular dependence of the intensity of the light transmitted through
the film~\cite{4}.  Perhaps more interesting was the discovery of
satellite peaks in the angular dependence of the intensity of the
light scattered from or transmitted through the film.  These are
well-defined peaks present on both sides of the enhanced
backscattering and enhanced transmission peaks, respectively, that
arise from the coherent interference of guided waves
with the frequency of the incident light, but with different
wavenumbers~\cite{5}.

It should be noted, however, that the prediction of these satellite peaks was first made in the context of the scattering of electromagnetic waves from a dielectric film containing a random distribution of volume scatterers ~\cite{Freilikher1994467}, rather than from a randomly rough surface, when the thickness of the film is small compared to the mean free path of the electromagnetic wave in the random medium.

In analytic~\cite{6,7,8} and computer simulation
calculations~\cite{9,10,15} of the multiple scattering of light from
clean two-dimensional randomly rough surfaces of perfect conductors
and penetrable media, enhanced backscattering was observed in the
results.  However, when attention turned to the scattering of light
from a perfectly conducting surface coated with a dielectric film,
conflicting results were obtained.  In these studies the
dielectric-perfect conductor interface was assumed to be planar, while
the vacuum-dielectric interface was assumed to be a two-dimensional
randomly rough interface.  In the first of these studies Kawanishi
\textit{et al}.~\cite{11} applied the stochastic functional approach to
this problem and found no evidence for satellite peaks in their
results.  They suggested that the ensemble averaging of the intensity
of the scattered field restores isotropy in the mean scattering plane,
and thereby eliminates the occurrence of special scattering angles at
which satellite peaks could occur.  In subsequent work in which the
reduced Rayleigh equation for scattering from this structure~\cite{8,Soubret:01}
was solved in the form of expansions of the amplitudes of the p- and
s-polarized components of the scattered field in powers of the surface
profile function through terms of third order, satellite peaks were
found.  However, the contribution to the scattering amplitudes
associated with the third-order term was larger than that from the
first-order term for the roughness and experimental parameters assumed
in that work.  It is therefore possible that these values fell outside
the ranges for which a perturbative solution of the reduced Rayleigh
equation is reliable.

Although satellite peaks were observed in experiments carried out by Méndez \textit{et al.}~\cite{Mendez:99} that utilized the double passage of polarized light through a random phase screen, the experimental conditions were sufficiently different from those studied theoretically in Refs.~\cite{8,11}, that these results could not be used to support the predictions of either of these studies.

In an effort to resolve the issue of whether satellite peaks do or do
not exist in the scattering of light from a rough dielectric film
deposited on the planar surface of a metal, in this paper we carry out
a  nonperturbative, purely numerical, solution of the
reduced Rayleigh equation~\cite{Rayleigh2DImplementation} for the scattering of p- and
s-polarized light from a structure consisting of a dielectric
film deposited on a metal substrate when the dielectric-metal
interface is planar, 
while the vacuum-dielectric interface is a two-dimensional
randomly rough interface.  This is an approach that was used
successfully in recent calculations of the scattering of p-
and s-polarized light from a two-dimensional randomly
rough interface between a  dielectric and a metal~\cite{Rayleigh2DImplementation,12},
which prompts its application to the present problem.

The system we study consists of vacuum $(\varepsilon_1)$ in the region
$x_3 > d + \zeta\left(\pvec{x}\right)$, where $\pvec{x} = (x_1,x_2,0)$; a
dielectric film $(\varepsilon_2)$ in the region $0 < x_3 < d +
\zeta\left(\pvec{x}\right)$; and a lossy metal $(\varepsilon_3)$ in the region
$x_3 < 0$.  The surface profile function $\zeta\left(\pvec{x}\right)$ is assumed
to be a single-valued function of $\pvec{x}$ that is differentiable
with respect to $x_1$ and $x_2$, and constitutes a zero-mean,
stationary, isotropic, Gaussian random process defined by 
\begin{align}
  \left<\zeta(\pvec{x}) \zeta(\pvec{x}') \right> &=
        \delta^2   W\left( \left| \pvec{x} - \pvec{x}' \right| \right). 
\end{align}
The angle brackets here denote an average over the ensemble of
realizations of the surface profile function, and 
$\delta = \left\langle \zeta^2\left(\pvec{x} \right) \right\rangle^{1/2}$ 
is the rms height of the surface roughness.

The electric field in the vacuum $\left[x_3 > d+\zeta\left(\pvec{x}\right)\right]$ is the
sum of an incident field and a scattered field, $\vec{E}(\vec{x}; t) =
\left[\vec{E}(\vec{x} |\omega )_{\mathrm{inc}}+\vec{E}(\vec{x} |\omega )_{\mathrm{sc}}\right] \exp \left( -\imu\omega t \right)$,
where 
\begin{subequations}
  \begin{align}
  \begin{aligned}
    \vec{E}(\vec{x} |\omega )_{\mathrm{inc}} 
        &= \bigg\{ \frac{c}{\omega} \left[ \vec{\hat{k}}_\parallel \alpha_1(k_\parallel ) 
                                           + \vec{\hat{x}}_3 k_\parallel \right] B_\mathrm{p}(\pvec{k} )
            + \left(\vec{\hat{x}}_3 \times \vec{\hat{k}}_\parallel \right) B_\mathrm{s}(\pvec{k} )\bigg\} \\
        &\times\exp \left(\imu \pvec{k}\cdot\pvec{x}  - \imu \alpha_1(k_\parallel )x_3\right)
  \end{aligned}
  \end{align}
  \begin{align}
  \begin{aligned}
    \vec{E}(\vec{x} |\omega )_{\mathrm{sc}} 
    &= \int \!\frac{\dint^2 q_\parallel}{(2\pi )^2} 
       \bigg\{ \frac{c}{\omega} 
        \left[ \vec{\hat{q}}_\parallel \alpha_1(q_\parallel )
              -\vec{\hat{x}}_3 q_\parallel \right] A_\mathrm{p}(\pvec{q} )
        + \left(\vec{\hat{x}}_3 \times \pvec{\hat{q}} \right)  A_\mathrm{s}(\pvec{q})
       \bigg\}  \\
       &\times \exp \left( \imu\pvec{q}\cdot\pvec{x} + \imu\alpha_1(q_\parallel )x_3\right),
  \end{aligned}
  \end{align}
\end{subequations}
while the subscripts p and s denote the
p-polarized and s-polarized components of these
fields with respect to the local planes of incidence and scattering.
A caret over a vector indicates that it is a unit vector, and the vector $\mathbf{k}_\parallel$ is defined as $\mathbf{k}_\parallel = \left( k_1, k_2, 0 \right)$ (with similar definitions for $\mathbf{q}_\parallel$ and $\mathbf{p}_\parallel$). The
functions $\alpha_i(q_\parallel )$ $(i = 1, 2, 3)$ are defined by
\begin{align}
  \alpha_i(q_\parallel) =
    \left[
        \varepsilon_i \left( \frac{\omega}{c} \right)^2 - q_\parallel^{2}
    \right]^{1/2}, 
  \qquad \Re\,\alpha_i(q_\parallel) > 0,
  \, \Im\,\alpha_i(q_\parallel) > 0 . 
  \label{eq:3} 
\end{align}

A linear relation exists between the amplitudes $A_{\alpha}(\pvec{q})$
and $B_{\beta}(\pvec{k})$ $(\alpha,\beta =
\mathrm{p}, \mathrm{s})$, which we write as 
\begin{align}
  A_{\alpha}(\pvec{q} ) &= 
  \sum_{\beta} R_{\alpha\beta}(\pvec{q} |\pvec{k} ) B_{\beta}(\pvec{k})
\end{align}
where $R_{\alpha\beta}$ is the scattering amplitude for incident $\beta$-polarized light scattered into $\alpha$-polarized light.
The convention we use with respect to the polarization subscripts is
\begin{align}
    \vec{R}(\pvec{q} |\pvec{k}) =
    \begin{pmatrix}
        R_{\mathrm{pp}}(\pvec{q} |\pvec{k})
            & R_{\mathrm{ps}}(\pvec{q} |\pvec{k}) \\
        R_{\mathrm{sp}}(\pvec{q} |\pvec{k})
            & R_{\mathrm{ss}}(\pvec{q} |\pvec{k})
    \end{pmatrix}.
\end{align}

It has been shown by Soubret \textit{et al}.~\cite{8} and
Leskova~\cite{13} that the scattering amplitudes $\left[
R_{\alpha\beta}(\pvec{q} |\pvec{k} ) \right]$ satisfy the matrix integral
equation
\begin{align}
  \int \frac{\dint^2 q_\parallel}{(2\pi )^2} 
     \vec{M}(\pvec{p} |\pvec{q}) 
       \vec{R}(\pvec{q} |\pvec{k}) 
      &= - \vec{N}(\pvec{p} |\pvec{k} ) , 
  \label{eq:6}
\end{align}
called a \textit{reduced Rayleigh equation} because it is an equation for
only the scattered field in the medium of incidence, and not for the
fields in the film and in the substrate.The effects of the latter
two fields are contained in the elements of the matrices
$\vec{M}(\pvec{p} |\pvec{q})$ and $\vec{N}(\pvec{p} |\pvec{k})$. With the shorthand notation $\alpha(q_\parallel, \omega) \equiv \alpha(q_\parallel)$, the elements of these matrices in the forms obtained by Leskova~\cite{13}
are 
\begin{subequations}
    \label{eq:7}
\begin{align}
    %
    %
&\begin{aligned}
  M_{\mathrm{pp}}&(\pvec{p} | \pvec{q})
  =
  \left[
      p_\parallel q_\parallel 
    + \alpha_2(p_\parallel) ( \pvec{\hat{p}}\cdot\pvec{\hat{q}} ) \alpha_1(q_\parallel) 
  \right] \\
  &\times\Gamma_\mathrm{p}(p_\parallel) \exp\left(-\imu\left[ \alpha_2(p_\parallel)-\alpha_1(q_\parallel)\right]d \right)
  \frac{ 
         I\left(\alpha_2(p_\parallel)-\alpha_1(q_\parallel) | \pvec{p} - \pvec{q} \right)
       }{
        \alpha_2(p_\parallel)-\alpha_1(q_\parallel)
        }
   \\ 
   & 
    +
  \left[
      p_\parallel q_\parallel 
    - \alpha_2(p_\parallel) ( \pvec{\hat{p}}\cdot\pvec{\hat{q}} ) \alpha_1(q_\parallel) 
  \right] \\
  &\times\Delta_\mathrm{p}(p_\parallel) \exp\left(\imu\left[ \alpha_2(p_\parallel)+\alpha_1(q_\parallel)\right]d \right)
  \frac{ 
         I\left(-\left[\alpha_2(p_\parallel)+\alpha_1(q_\parallel)\right] | \pvec{p} - \pvec{q} \right)
       }{
        \alpha_2(p_\parallel)+\alpha_1(q_\parallel)
        }
    \label{eq:7a}
\end{aligned}
\\
%
%
%
%
    %
    %
&\begin{aligned}
  M_{\mathrm{ps}}&(\pvec{p}  |\pvec{q}) 
  =
   - \frac{\omega}{c} \alpha_2(p_\parallel) \left( \pvec{\hat{p}}\times\pvec{\hat{q}} \right)_3
   \\
   &\left(
  \Gamma_\mathrm{p}(p_\parallel) \exp\left(-\imu\left[ \alpha_2(p_\parallel)-\alpha_1(q_\parallel)\right]d \right)
  \frac{ 
         I\left(\alpha_2(p_\parallel)-\alpha_1(q_\parallel) | \pvec{p} - \pvec{q} \right)
       }{
        \alpha_2(p_\parallel)-\alpha_1(q_\parallel)
        }
   \right.
   \\ 
   & 
   \qquad 
    -
  \left.
  \Delta_\mathrm{p}(p_\parallel) \exp\left(\imu\left[ \alpha_2(p_\parallel)+\alpha_1(q_\parallel)\right]d \right)
  \frac{ 
         I\left(-\left[\alpha_2(p_\parallel)+\alpha_1(q_\parallel)\right] | \pvec{p} - \pvec{q} \right)
       }{
        \alpha_2(p_\parallel)+\alpha_1(q_\parallel)
        }
  \right)
    \label{eq:7b}
\end{aligned}
\\
%
%
%
%
    %
    %
&\begin{aligned}
  M_{\mathrm{sp}}&(\pvec{p}  |\pvec{q}) 
  =
    \frac{\omega}{c} \left(\pvec{\hat{p}}\times\pvec{\hat{q}}\right)_3   \alpha_1(q_\parallel)  \\
   &\left(
  \Gamma_\mathrm{s}(p_\parallel) \exp
    \left(
        -\imu\left[\alpha_2(p_\parallel)-\alpha_1(q_\parallel)\right]d
    \right)
  \frac{ 
         I\left(\alpha_2(p_\parallel)-\alpha_1(q_\parallel) | \pvec{p} - \pvec{q} \right)
       }{
        \alpha_2(p_\parallel)-\alpha_1(q_\parallel)
        }
   \right.
   \\ 
   & 
   \qquad 
    +
  \left.
  \Delta_\mathrm{s}(p_\parallel) \exp\left(\imu\left[\alpha_2(p_\parallel)+\alpha_1(q_\parallel)\right]d\right)
  \frac{ 
         I\left(-\left[\alpha_2(p_\parallel)+\alpha_1(q_\parallel)\right] | \pvec{p} - \pvec{q} \right)
       }{
        \alpha_2(p_\parallel)+\alpha_1(q_\parallel)
        }
  \right)
    \label{eq:7c}
\end{aligned}
\\
%
%
%
%
    %
    %
&\begin{aligned}
  M_{\mathrm{ss}}&(\pvec{p}  |\pvec{q})
  =
  \frac{\omega^2}{c^2} \left( \pvec{\hat{p}}\cdot\pvec{\hat{q}} \right)  \\
  &\left(
  \Gamma_\mathrm{s}(p_\parallel) \exp
    \left(        -\imu\left[ \alpha_2(p_\parallel)-\alpha_1(q_\parallel)\right]d 
    \right)
  \frac{ 
         I\left(\alpha_2(p_\parallel)-\alpha_1(q_\parallel) | \pvec{p} - \pvec{q} \right)
       }{
        \alpha_2(p_\parallel)-\alpha_1(q_\parallel)
        }
   \right.
   \\ 
   & 
   \qquad 
    +
  \left.
  \Delta_\mathrm{s}(p_\parallel) \exp
    \left(
        \imu\left[ \alpha_2(p_\parallel)+\alpha_1(q_\parallel)\right]d 
    \right)
  \frac{ 
         I\left(
            -\left[\alpha_2(p_\parallel)+\alpha_1(q_\parallel)\right]
                | \pvec{p} - \pvec{q}
            \right)
       }{
        \alpha_2(p_\parallel)+\alpha_1(q_\parallel)
        }
  \right),
    \label{eq:7d}
\end{aligned}
\end{align}
\end{subequations}
and
\begin{subequations}
\label{eq:8}
\begin{align}
    %
    %
&\begin{aligned}
  N_{\mathrm{pp}}&(\pvec{p} | \pvec{k})
  =
  -\left[
      p_\parallel k_\parallel 
    - \alpha_2(p_\parallel) ( \pvec{\hat{p}}\cdot\pvec{\hat{k}} ) \alpha_1(k_\parallel) 
  \right] \\
  &\times\Gamma_\mathrm{p}(p_\parallel) \exp
    \left(
        -\imu\left[ \alpha_2(p_\parallel)+\alpha_1(k_\parallel)\right]d 
    \right)
  \frac{ 
         I\left(\alpha_2(p_\parallel)+\alpha_1(k_\parallel) | \pvec{p} - \pvec{k} \right)
       }{
        \alpha_2(p_\parallel)+\alpha_1(k_\parallel)
        }
   \\ 
   & -
  \left[
      p_\parallel k_\parallel 
    + \alpha_2(p_\parallel) ( \pvec{\hat{p}}\cdot\pvec{\hat{k}} ) \alpha_1(k_\parallel) 
  \right] \\
  &\times\Delta_\mathrm{p}(p_\parallel) \exp
  \left(
    \imu\left[ \alpha_2(p_\parallel)-\alpha_1(k_\parallel)\right]d
  \right)
  \frac{ 
         I\left(
            -\left[\alpha_2(p_\parallel)-\alpha_1(k_\parallel)\right]
                | \pvec{p} - \pvec{k}
            \right)
       }{
        \alpha_2(p_\parallel)-\alpha_1(k_\parallel)
        }
\label{eq:8a}
\end{aligned}
\\
%
%
%
%
    %
    %
&\begin{aligned}
  N_{\mathrm{ps}}&(\pvec{p}  |\pvec{k}) 
  =
   -\frac{\omega}{c} \alpha_2(p_\parallel) \left( \pvec{\hat{p}}\times\pvec{\hat{k}} \right)_3 \\
  &\times\left(
  \Gamma_\mathrm{p}(p_\parallel) \exp
    \left(
        -\imu\left[ \alpha_2(p_\parallel)+\alpha_1(k_\parallel)\right]d 
    \right)
  \frac{ 
        I\left(
            \alpha_2(p_\parallel)+\alpha_1(k_\parallel)
                | \pvec{p} - \pvec{k}
        \right)
       }{
        \alpha_2(p_\parallel)+\alpha_1(k_\parallel)
        }
   \right.
   \\ 
   &- 
  \left.
  \Delta_\mathrm{p}(p_\parallel) \exp
    \left(
        \imu\left[ \alpha_2(p_\parallel)-\alpha_1(k_\parallel)\right]d 
    \right)
  \frac{ 
         I\left(
            -\left[\alpha_2(p_\parallel)-\alpha_1(k_\parallel)\right]
                | \pvec{p} - \pvec{k}
        \right)
       }{
        \alpha_2(p_\parallel)-\alpha_1(k_\parallel)
        }
  \right)
\label{eq:8b}
\end{aligned}
\\
%
%
%
%
    %
    %
&\begin{aligned}
  N_{\mathrm{sp}}&(\pvec{p}  |\pvec{k}) 
  =
    \frac{\omega}{c} \left(\pvec{\hat{p}}\times\pvec{\hat{k}}\right)_3   \alpha_1(k_\parallel) \\
   &\times\left(
  \Gamma_\mathrm{s}(p_\parallel) \exp
    \left(
        -\imu\left[\alpha_2(p_\parallel)+\alpha_1(k_\parallel)\right]d
    \right)
  \frac{ 
        I\left(
            \alpha_2(p_\parallel)+\alpha_1(k_\parallel)
                | \pvec{p} - \pvec{k}
        \right)
       }{
        \alpha_2(p_\parallel)+\alpha_1(k_\parallel)
        }
   \right.
   \\ 
   & +
  \left.
  \Delta_\mathrm{s}(p_\parallel) \exp
    \left(
        \imu\left[\alpha_2(p_\parallel)-\alpha_1(k_\parallel)\right]d
    \right)
  \frac{ 
        I\left(
            -\left[\alpha_2(p_\parallel)-\alpha_1(k_\parallel)\right]
                | \pvec{p} - \pvec{k}
        \right)
       }{
        \alpha_2(p_\parallel)-\alpha_1(k_\parallel)
        }
  \right)
\label{eq:8c}
\end{aligned}
\\
%
%
%
%
    %
    %
&\begin{aligned}
  N_{\mathrm{ss}}&(\pvec{p}  |\pvec{k})
  =
 \frac{\omega^2}{c^2} \left( \pvec{\hat{p}}\cdot\pvec{\hat{k}} \right) \\
  & \times \left(
  \Gamma_s(p_\parallel) \exp
    \left(
        -\imu\left[ \alpha_2(p_\parallel)+\alpha_1(k_\parallel)\right]d 
    \right)
  \frac{ 
        I\left(
            \alpha_2(p_\parallel)+\alpha_1(k_\parallel)
            | \pvec{p} - \pvec{k}
        \right)
       }{
        \alpha_2(p_\parallel)+\alpha_1(k_\parallel)
        }
   \right.
   \\ 
   & +
  \left.
  \Delta_s(p_\parallel) \exp
    \left(
        \imu\left[ \alpha_2(p_\parallel)-\alpha_1(k_\parallel)\right]d 
    \right)
  \frac{ 
         I\left(
            -\left[\alpha_2(p_\parallel)-\alpha_1(k_\parallel)\right]
            | \pvec{p} - \pvec{k}
         \right)
       }{
        \alpha_2(p_\parallel)-\alpha_1(k_\parallel)
        }
  \right).
\label{eq:8d}
\end{aligned}
\end{align}
\end{subequations}

In writing Eqs.~\eqref{eq:7} and \eqref{eq:8} we have introduced the functions
\begin{subequations}
  \label{eq:9}
  \begin{align}
    \Gamma_\mathrm{p}(p_\parallel ) &= 
    \varepsilon_2\alpha_3(p_\parallel, \omega) 
    + \varepsilon_3\alpha_2(p_\parallel, \omega) 
    \label{eq:9a}\\
    \Delta_\mathrm{p}(p_\parallel ) &= 
    \varepsilon_2\alpha_3(p_\parallel, \omega) -
    \varepsilon_3\alpha_2(p_\parallel, \omega) 
    \label{eq:9b}
  \end{align}
\end{subequations}
and
\begin{subequations}
  \label{eq:10}
  \begin{align}
    \Gamma_\mathrm{s}(p_\parallel ) &= 
    \alpha_3(p_\parallel, \omega) + \alpha_2(p_\parallel, \omega) 
    \label{eq:10a}\\
    \Delta_\mathrm{s}(p_\parallel ) &= 
    \alpha_3(p_\parallel, \omega) - \alpha_2(p_\parallel, \omega) ,
    \label{eq:10b}
  \end{align}
\end{subequations}
as well as
\begin{align}
  I \left( \gamma |\vec{Q}_\parallel \right) =& 
  \int\!\dint^2x_\parallel\, 
   \exp \left(-\imu\pvec{Q} \cdot \pvec{x} \right) \exp \left[-\imu\gamma \zeta \left( \pvec{x} \right) \right] .
   \label{eq:11}
\end{align}
The scattering amplitudes $\left[ R_{\alpha\beta}(\pvec{q} |\pvec{k}) \right]$
play a central role in the present theory because the mean
differential reflection coefficient, an experimentally measurable
function, can be expressed in terms of these amplitudes.  The
differential reflection coefficient $(\partial R/\partial \Omega_s)$
is defined such that $(\partial R/\partial\Omega_s)\dint\Omega_s$ is the
fraction of the total time-averaged flux incident on the surface that
is scattered into the element of solid angle $\dint\Omega_s$ about the
scattering direction $(\theta_s ,\phi_s)$.  Since we are studying the
scattering of light from a randomly rough surface, it is the average
of this function over the ensemble of realizations of the surface
profile function that we need to calculate.  The contribution to the
mean differential reflection coefficient from the incoherent (diffuse)
component of the scattered light, when incident light of $\beta$
polarization whose wave vector has the projection $\pvec{k}$ on the
mean scattering surface is scattered into light of $\alpha$
polarization whose wave vector has the projection $\pvec{q}$ on the
mean scattering surface, denoted $\left\langle \partial R_{\alpha\beta} / \partial\Omega_s \right\rangle_{\mathrm{incoh}}$, is given by
\begin{subequations}
\label{eq:12}
\begin{align}
  \left< \frac{\partial R_{\mathrm{pp}}}{\partial\Omega_s}\right>_{\mathrm{incoh}} 
  =&  \frac{1}{S} 
      \frac{\sqrt{\varepsilon_1}}{4\pi^2} 
      \frac{\omega}{c} 
      \frac{\alpha^2_1(q_\parallel)}{\alpha_1(k_\parallel)} 
      \left[ 
              \left< \left|  R_{\mathrm{pp}}(\pvec{q} | \pvec{k}) \right|^2\right> 
           -  \left| \left<  R_{\mathrm{pp}}(\pvec{q} | \pvec{k}) \right> \right|^2 
      \right]
  \label{eq:12a}\\
  \left< \frac{\partial R_{\mathrm{ps}}}{\partial\Omega_s}\right>_{\mathrm{incoh}} 
  =&  \frac{1}{S} 
      \frac{\varepsilon_1^{3/2}}{4\pi^2} 
      \frac{\omega}{c} 
      \frac{\alpha^2_1(q_\parallel)}{\alpha_1(k_\parallel)} 
      \left[ 
              \left< \left|  R_{\mathrm{ps}}(\pvec{q} | \pvec{k}) \right|^2\right> 
           -  \left| \left<  R_{\mathrm{ps}}(\pvec{q} | \pvec{k}) \right> \right|^2 
      \right]
  \label{eq:12b}\\
  \left< \frac{\partial R_{\mathrm{sp}}}{\partial\Omega_s}\right>_{\mathrm{incoh}} 
  =&  \frac{1}{S} 
      \frac{1}{4\pi^2\sqrt{\varepsilon_1}} 
      \frac{\omega}{c} 
      \frac{\alpha^2_1(q_\parallel)}{\alpha_1(k_\parallel)} 
      \left[ 
              \left< \left|  R_{\mathrm{sp}}(\pvec{q} | \pvec{k}) \right|^2\right> 
           -  \left| \left<  R_{\mathrm{sp}}(\pvec{q} | \pvec{k}) \right> \right|^2 
      \right]
  \label{eq:12c}\\
  \left< \frac{\partial R_{\mathrm{ss}}}{\partial\Omega_s}\right>_{\mathrm{incoh}} 
  =&  \frac{1}{S} 
      \frac{\sqrt{\varepsilon_1}}{4\pi^2} 
      \frac{\omega}{c} 
      \frac{\alpha^2_1(q_\parallel)}{\alpha_1(k_\parallel)} 
      \left[ 
              \left< \left|  R_{\mathrm{ss}}(\pvec{q} | \pvec{k}) \right|^2\right> 
           -  \left| \left<  R_{\mathrm{ss}}(\pvec{q} | \pvec{k}) \right> \right|^2 
      \right],
  \label{eq:12d}
\end{align}
\end{subequations}
where $S$ is the area of the plane $x_3 = 0$ covered by the rough
surface.  The two-dimensional wave vectors $\pvec{k}$ and $\pvec{q}$
are defined in terms of the polar and azimuthal angles of incidence
$(\theta_0,\phi_0)$ and scattering $(\theta_s,\phi_s)$, respectively,
by $\pvec{k} = \sqrt{\varepsilon_1}(\omega /c)$ $\sin\theta_0$
$(\cos\phi_0, \sin\phi_0, 0)$ and $\pvec{q} = \sqrt{\varepsilon_1} (\omega
/c)\sin\theta_s(\cos \phi_s, \sin\phi_s, 0)$.  Thus these wave vectors
in Eq.~\eqref{eq:12} are restricted to the domains $k_\parallel <
\sqrt{\varepsilon_1} (\omega /c)$ and $q_\parallel <
\sqrt{\varepsilon_1} (\omega /c)$ of the $q_1q_2$ plane.

Up to now Eq.~\eqref{eq:6} has been solved by small-amplitude
perturbation theory through terms of third order in the surface
profile function~\cite{8,Johnson:99}.  Here we present results for the mean
differential reflection coefficient and for the full angular
distribution of the intensity of the scattered light obtained by a
nonperturbative, purely numerical solution of
Eqs.~\eqref{eq:6}--\eqref{eq:11}, as described in Ref.~\cite{Rayleigh2DImplementation}.  This was done by generating a
realization of the surface profile function numerically on a grid of
$N^2_x$ points within a square region of the $x_1x_2$ plane of edge
$L$, so that the (linear) sampling interval was $\Delta x = L/N_x$.  A
two-dimensional version of the filtering method used in~\cite{14,Rayleigh2DImplementation}  was used to generate the profile
function~\cite{15}.  The function $I(\gamma |\pvec{Q} )$ was then
evaluated by expanding the integrand in powers of the surface profile
function $\zeta(\pvec{x})$, and calculating the Fourier transform of
$\zeta^n(\pvec{x})$ by the fast Fourier transform.  In evaluating the
integral over $\pvec{q}$ in Eq.~\eqref{eq:6} the infinite limits were replaced
by finite ones: $\left(q_1^2  + q_2^2\right)^{1/2} \leq Q/2$.  The Nyquist sampling
theorem requires that $|q_1|$ and $|q_2|$ be smaller than $Q_c = \pi
/\Delta x$~\cite[p. 605]{17}. The components of the vector $\pvec{p} -
\pvec{q}$ entering $I(\gamma |\pvec{p} - \pvec{q} )$ lie in the
interval $[-Q,Q]$, so we have chosen $Q = Q_c$.  A grid with a
grid constant $\Delta q_1 = \Delta q_2 = \Delta q = 2\pi /L$ was constructed within the circular
region of the $q_1q_2$ plane where $\left(q_1^2 + q_2^2\right) \leq Q/2$.
The integral over this
region in Eq.~\eqref{eq:6} was carried out by a two-dimensional version of the
extended midpoint method~\cite[p. 161]{17} and the values of
$R_{\alpha\beta}(\pvec{q} |\pvec{k} )$ were calculated for values of
$\pvec{q}$ at the points of this grid for a given value of $\pvec{k}$,
which was also a point on this grid.  The resulting matrix equations
were solved by LU factorization and backsubstitution.  The values of
$R_{\alpha\beta}(\pvec{q} |\pvec{k} )$ and $|R_{\alpha\beta}(\pvec{q}
|\pvec{k} )|^2$ were then calculated for $N_p$ realizations of the
surface profile function.  An arithmetic average of the $N_p$ results
for each of these functions yielded the averages $\left\langle
R_{\alpha\beta}(\pvec{q} |\pvec{k})\right\rangle$ and $\left\langle
|R_{\alpha\beta}(\pvec{q} |\pvec{k})|^2 \right\rangle$, from which the incoherent contribution to the mean differential reflection coefficients were calculated according to Eq.~\eqref{eq:12}.

We apply this approach to the scattering of p- and s-polarized plane waves, whose wavelength is $\lambda = 633$ nm, incident from vacuum $(\epsilon_1 = 1)$ on a dielectric film $(\epsilon_2 = 2.6896+0.01\mathrm{i})$ coating a silver surface $(\epsilon_3 = - 18.28 + 0.481\mathrm{i})$ \cite{19}.  The mean thickness of the film is $d = 0.756\lambda = 478.5$ nm.  The roughness of the vacuum-dielectric interface is characterized by a two-dimensional version of the West--O'Donnell power spectrum \cite{20} given by \cite{6}
\begin{align}
    g(|\pvec{k}|) = \frac{4\pi}{k^2_+-k^2_-} \theta (|\pvec{k}|-k_-)\theta (k_+-|\pvec{k}|),
    \label{eq:powerspectrum}
\end{align}
where $\theta (x)$ is the Heaviside unit step function, and $k_- = 0.82 (\omega /c)$, $k_+ = 1.97 (\omega /c)$.  The rms height of the surface roughness was assumed to be $\delta = \lambda /40 = 15.82$ nm, the surface was discretized on a grid of resolution $\Delta x_1 = \Delta x_2 = 0.123 \lambda = 77.6$ nm and the edge of the (quadratic) surface was $L = 55 \lambda = \unit{34.8}{\micro\meter}$.

The contribution to the mean differential reflection coefficient $\left\langle \partial R_{\alpha\beta}(\pvec{q}|\pvec{k} )/\partial\Omega_s \right\rangle_{\mathrm{incoh}}$ from single-scattering processes \big[second order in $\zeta (\pvec{x})$\big] is proportional to $g(|\pvec{q}-\pvec{k}|)$ \cite{6}.  Since the power spectrum \eqref{eq:powerspectrum} is identically zero for $|\pvec{k}| < k_-$, there is no contribution to the mean differential reflection coefficient from the light scattered incoherently by single-scattering processes when the wave vectors $\pvec{q}$ and $\pvec{k}$ satisfy the inequality $|\pvec{q}-\pvec{k}| < k_-$.  The contribution to $\left\langle \partial R_{\alpha\beta}(\pvec{q}|\pvec{k})/\partial \Omega_s\right\rangle_{\mathrm{incoh}}$ when this condition is satisfied is due only to multiple-scattering processes, including the enhanced backscattering peak and the satellite peaks.  These features are more clearly visible in this case because they do not ride on a large background due to single-scattering processes.  This is the reason that the calculations whose results are presented here were carried out on the basis of the power spectrum~\eqref{eq:powerspectrum}.

In Fig. 1(a) we present the contribution to the mean differential reflection coefficient from the light scattered incoherently as functions of the polar scattering angle $\theta_s$ for the in-plane $(\phi_s = \phi_0 = 45^{\circ}$) co-(p$\to$p, s$\to$s) and cross-(p$\to$s, s$\to$p) polarized scattering when a p- or s-polarized plane wave is incident on the dielectric surface at angles of incidence $(\theta_0, \phi_0)$ given by ($0.74^{\circ}, 45^{\circ})$. (In figures showing in-plane or out-of-plane scattering, we depart from the commonly accepted principle of not using negative polar angles, in that we allow for negative $\theta_s$.).  An arithmetic average of results obtained for $N_p = 11,165$ realizations of the surface profile function was carried out to produce these figures.  In Fig. 1(b) we present the analogous results for out-of-plane $(\phi_s=\phi_0\pm 90^{\circ})$ scattering when the roughness and experimental parameters have the values assumed in generating Fig. 1(a).
%
%
\begin{figure}[htbp]
    \centering
    \subfigure{
        \includegraphics{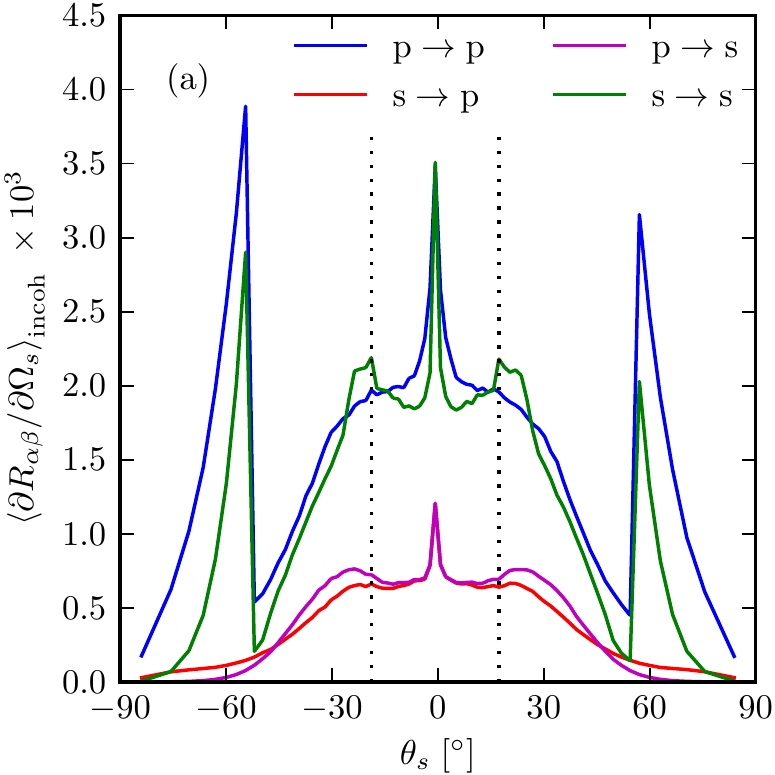}
    }
    \subfigure{
        \includegraphics{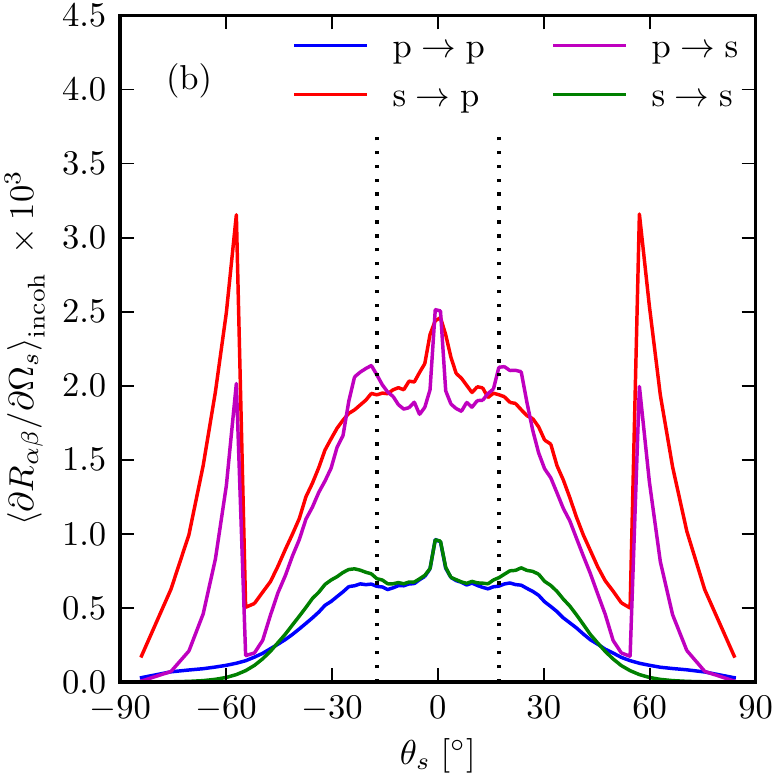}
    }
    \caption{\label{fig:normal-incidence-cuts}
        (a)  The contributions to the mean differential reflection coefficient as functions of the polar scattering angle $\theta_s$ from the in-plane $(\phi_s = \phi_0)$ co-polarized (p$\to$p, s$\to$s) and cross-polarized  (p$\to$s, s$\to$p) scattering of light incident on the two-dimensional randomly rough surface of a dielectric film deposited on the planar surface of silver, whose dielectric constant is $\epsilon_3 = -18.28 + 0.481\mathrm{i}$.  The wavelength of the incident light is $\lambda = 633$ nm, the angles of incidence are $(\theta_0,\phi_0) = (0.74^{\circ}, 45^{\circ})$.  The dielectric constant of the film is $\epsilon_2 = 2.6896 + 0.01\mathrm{i}$,  and its mean thickness is $d = 478.5$ nm.  The roughness of the surface is characterized by the power spectrum in Eq.~\eqref{eq:powerspectrum}, with $k_- = 0.82 (\omega /c )$, $k_+ = 1.97 (\omega /c)$, and its rms height is $\delta = \lambda / 40 = 15.82$ nm.  (b) The same as (a) for out-of-plane $(\phi_s = \phi_0 +90^{\circ})$ scattering.}
\end{figure}

In the results depicted in Fig. 1(a) [1(b)] single-scattering processes give no contributions to the mean differential reflection coefficient for $-53.8^{\circ} < \theta_s < 56.4^{\circ}$ $(-55.08^{\circ} < \theta_s < 55.08^{\circ})$.    In both figures a well-defined enhanced backscattering peak is seen in the retroreflection direction.  In addition, in Fig. 1(a) additional peaks are seen on both sides of the enhanced backscattering peak in the s $\to$ s co-polarized scattering contribution to the mean differential  reflection coefficient.  These peaks are identified as satellite peaks.

We base this identification on the following consideration.  It was shown in~\cite{5} that in the in-plane co-polarized scattering of light of frequency $\omega$ from a one-dimensional randomly rough surface of a film system when the plane of incidence is perpendicular to the generators of the surface, satellite peaks occur at scattering angles given by
\begin{equation}
    \sin\theta_s^{(m,n)} =
        - \sin\theta_0\pm
            \frac{c}{\omega\sqrt{\varepsilon_1}}
            \left[ q_m(\omega )-q_n(\omega ) \right].
    \label{eq:13}
\end{equation}
The wave numbers $q_1(\omega ), q_2(\omega ), \ldots, q_N(\omega )$ are the wavenumbers of the guided waves supported by the film structure at the frequency of the incident light.  Not all of the peaks predicted by Eq.~(\ref{eq:13}) may be present in the mean differential reflection coefficient.  This happens when the absolute value of the right-hand side of Eq.~(\ref{eq:13}) is greater than unity.   Then the corresponding peak lies in the nonradiative region of the $q_1 q_2$ plane.  In addition, among the real satellite peaks that should appear in the radiative region, not all may be sufficiently intense to be observable.

The scattering angles defined by Eq.~(\ref{eq:13}) are expected to give the angles at which satellite peaks occur in the in-plane co-polarized scattering from the two-dimensional randomly rough surface of the film system studied here.

In the absence of absorption and roughness the $\{ q_j(\omega )\}$ are the solutions of the dispersion relation
\begin{subequations}
\begin{align}
\begin{aligned}
    \alpha_2(q_\parallel, \omega)
    ={}& \frac{1}{2\epsilon_1\epsilon_3}
    \Big(
        \epsilon_2 \left[
            \epsilon_1\beta_3(q_\parallel, \omega) + \epsilon_3\beta_1(q_\parallel, \omega)
        \right]
        \cot\left[
            \alpha_2(q_\parallel, \omega)d
        \right] \\
        &\pm \big\{
            \epsilon^2_2 \left[
                \epsilon_1\beta_3(q_\parallel, \omega) + \epsilon_3\beta_1(q_\parallel, \omega)
            \right]^2
                \cot^2\left[\alpha_2(q_\parallel, \omega) d \right]
    \\
    & \qquad + 4\epsilon_1\epsilon^2_2\epsilon_3
        \beta_1(q_\parallel, \omega)\beta_3(q_\parallel,\omega)
        \big\}^{1/2}
    \Big)
\end{aligned}
\end{align}
in p polarization, and
\begin{align}
\begin{aligned}
    \alpha_2(q_\parallel, \omega)
    &= \frac{1}{2}
    \Big(
        \left[
            \beta_1(q_\parallel, \omega) + \beta_3(q_\parallel, \omega)
        \right]
        \cot\left[
            \alpha_2(q_\parallel, \omega) d
        \right] \\
        & \qquad \pm \big\{
            [\beta_1(q_\parallel, \omega)+\beta_3(q_\parallel, \omega)]^2
            \cot^2\left[
                \alpha_2(q_\parallel, \omega) d
            \right]
    \\
    & \qquad\qquad
        + 4 \beta_1(q_\parallel, \omega) + \beta_3(q_\parallel, \omega)
        \big\}^{1/2}
    \Big)
\end{aligned}
\end{align}
\end{subequations}
in s polarization.  In these equations $\beta_i(q_\parallel, \omega) = \left[ q_\parallel^2 -\epsilon_i(\omega/c)^2 \right]^{1/2}$ for $i = 1,3$, while $\alpha_{2}(q_\parallel, \omega)$ is defined in Eq.~\eqref{eq:3}.  The film structure studied in this paper is found  to support two guided waves in p-polarization, whose wave numbers are
\begin{subequations}
\label{eq:15}
\begin{align}
    q_1(\omega ) &= 1.4391 (\omega /c), \quad
    q_2(\omega ) = 1.0119 (\omega /c),
\end{align}
and two guided waves in s polarization, with wave numbers
\begin{align}
    q_1(\omega ) &= 1.5467 (\omega /c), \quad
    q_2(\omega ) =    1.2432 (\omega /c).
\end{align}
\end{subequations}
These results predict satellite peaks at scattering angles $\theta_s = -25.22^{\circ}$ and $23.74^{\circ}$ in p polarization and at $\theta_s = -18.13^{\circ}$ and $16.65^{\circ}$ in s polarization when we are considering in-plane scattering, assuming the same angles of incidence as in Fig.~\ref{fig:normal-incidence-cuts}.  These scattering angles are indicated by vertical dotted lines in Fig.~1(a).  The peaks at  $\theta_s = -18.13^{\circ}$ and $16.65^{\circ}$ are seen in the s$\to$s co-polarized scattering contribution to the mean differential reflection coefficient.  There is no evidence of satellite peaks at $\theta^{(1,2)}_s = -25.22^{\circ}$ and $23.74^{\circ}$ in the p$\to$p co-polarized scattering contribution to the mean differential reflection coefficient, presumably because they are too weak to be seen.  These results disagree with those of~\cite{8}, in which no satellite peaks were found in the in-plane s$\to$s scattering contribution to the mean differential reflection coefficient (although they are present in this contribution when the dielectric film is deposited on a planar perfectly conducting surface).  However,  in~\cite{8} the surface roughness was characterized by a Gaussian power spectrum, not the West--O'Donnell power spectrum assumed here.  The results of earlier calculations \cite{22} of the scattering of p- and s-polarized light from a film with a one-dimensional randomly rough surface characterized by a Gaussian power spectrum that is deposited on a planar perfectly conducting surface, display satellite peaks more weakly than when the roughness is characterized by a West--O'Donnell power spectrum~\cite{Simonsen_OptComm}.

Turning now to the results for out-of-plane scattering presented in Fig. 1(b), we see that an enhanced backscattering peak is present in each scattering configuration.   It is cut off in each configuration.  This is an artifact of the present calculation that results from the line defined by $\phi_s = \phi_0 \pm 90^{\circ}$ being exactly one grid point away from the backscattering direction.    It is important to note that in out-of-plane scattering the predominant contribution to the differential reflection coefficient is in the cross-polarized part. We see that the satellite peaks are now observed in the p$\to$s scattering configuration, meaning that incident p-polarized light excites both of the s-polarized guided modes with wave vectors in the $\phi = \phi_0 \pm 90^\circ$ directions, which subsequently interfere to cause satellite peaks in out-of-plane scattering. Hence, the well-known ``satellite peaks'' found in scattering from 1D surfaces turn into a kind of ``satellite rings'' for scattering from 2D surfaces, where part of the ring is co-polarized (s$\to$s in-plane) and part of the ring is cross-polarized (p$\to$s out-of-plane).
\begin{figure}[htbp]
    \centering
    \includegraphics{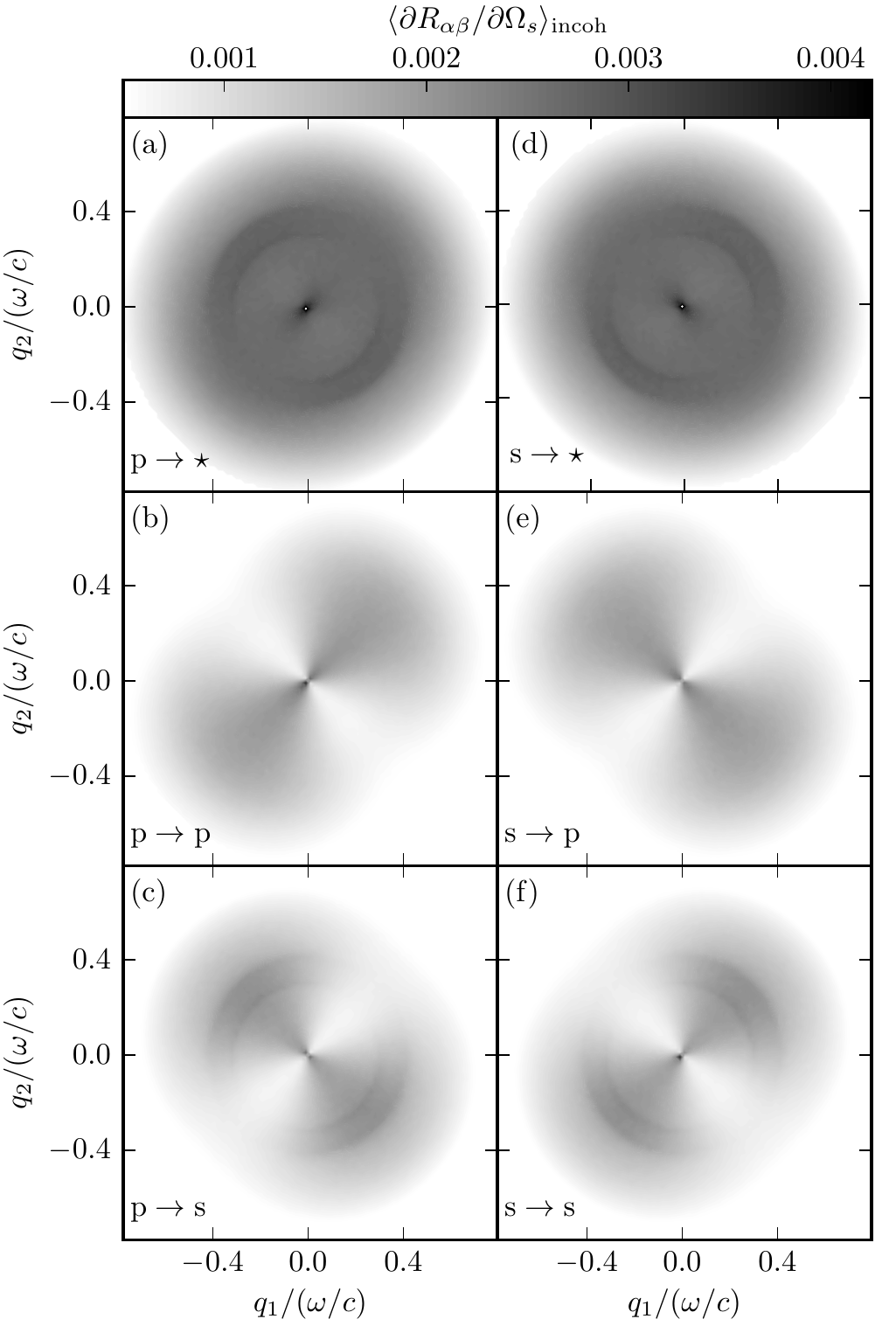}
    \caption{The complete angular distribution of the mean differential reflection coefficient $\langle \partial R_{\alpha\beta}/\partial\Omega_s\rangle_{\mathrm{incoh}}$ for the light scattered incoherently from the film structure. The material and experimental parameters assumed here are those used in obtaining the plots presented in Fig. 1. Light of either p (left column) or s (right column) polarization is incident on the structure.  In (a) and (d) all (diffusely) scattered light is recorded.  In (b) and (e) only the p-polarized scattered light is recorded, while in (c) and (f) only the s-polarized scattered light is recorded.  The dark dot in each panel indicates the enhanced backscattering peak. Note that the gray scale bar is cut at both ends in order to enhance the satellite rings. Also note that the contribution from single scattering is suppressed, i.e. the differential reflection coefficient is artificially set to $0$ for $\left|\pvec{q} - \pvec{k}\right| > k_-$.}
\end{figure}


In Fig. 2 we present contour plots of the complete angular distribution of the mean differential reflection coefficient for the light scattered incoherently from the film system studied here.  The material and experimental parameters used in producing these results have the same values used in obtaining Fig. 1.

\begin{figure}[!!tb]
    \centering
    \subfigure{
        \includegraphics{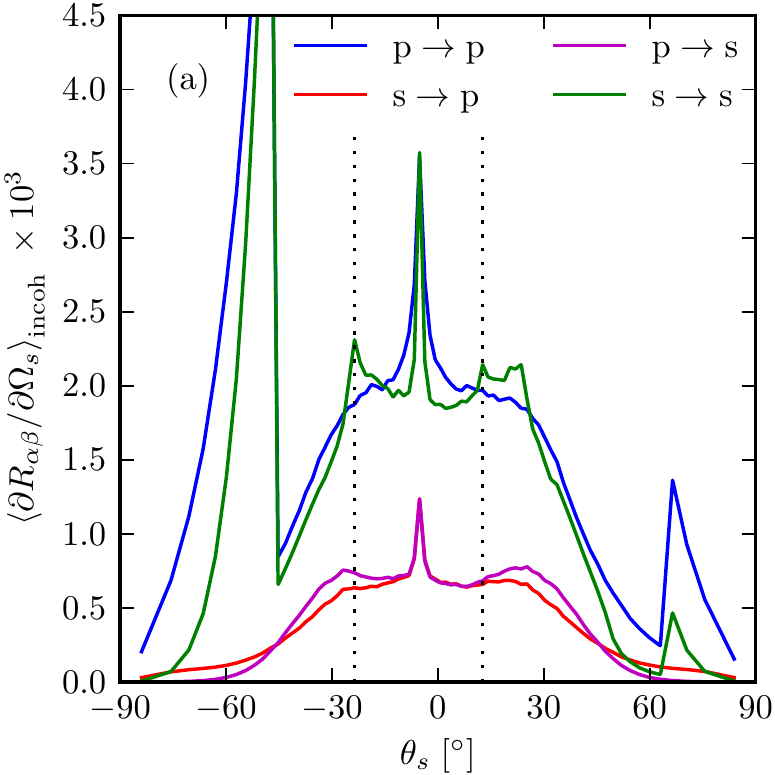}
    }
    \subfigure{
        \includegraphics{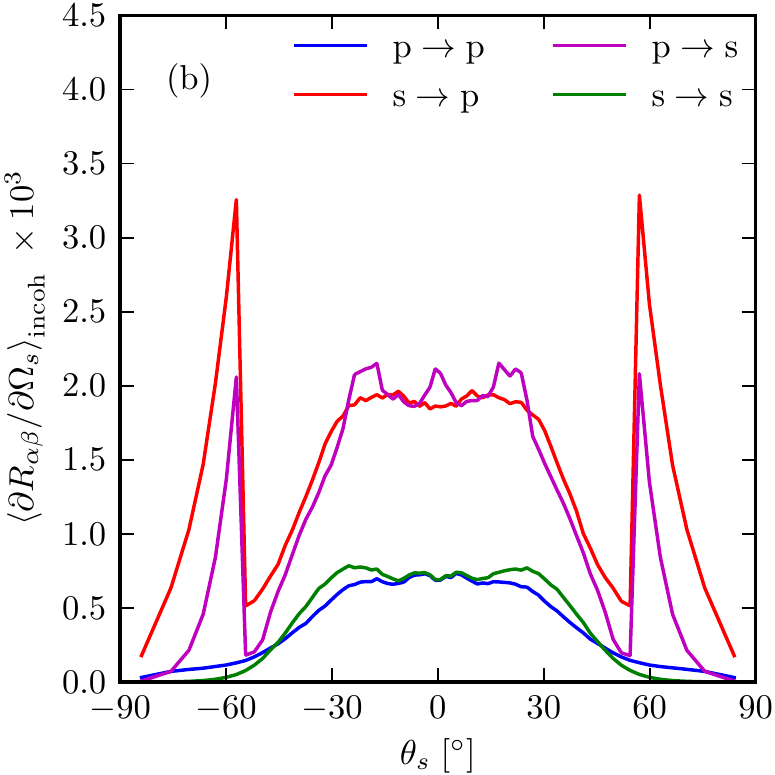}
    }
    \caption{
        The same as Fig. 1, but for angles of incidence given by $(\theta_0, \phi_0) = (5.19^{\circ}, 45^{\circ})$.}
\end{figure}

Light of p polarization (left column) or s polarization (right column) is incident on the structure.  In Figs. 2(a) and 2(d) all of the scattered light is recorded; in Figs. 2(b) and 2(e) only the p-polarized scattered light is recorded; while in Figs. 2(c) and 2(f) only the s-polarized scattered light is recorded.  In Fig. 2(f) we see two regions of high intensity in the in-plane polarized (s$\to$s) intensity distribution, centered at radii of approximately $0.29 (\omega  /c)$ at $\phi_s = 45^{\circ}$, and $0.31 (\omega /c)$ at $\phi_s = 225^{\circ}$.  These are the satellite peaks seen in the plot of $\left\langle \partial R_{\mathrm{ss}}/\partial\Omega_s \right\rangle_{\mathrm{incoh}}$ presented in Fig.~1(a).  No such regions of high intensity are seen in Fig. 2(b) at radii of $0.34 (\omega  /c)$ at $\phi_s = 45^{\circ}$ and $0.52 (\omega /c)$ at $\phi_s = 225^{\circ}$, where satellite peaks are predicted by Eq.~(\ref{eq:13}) for in-plane co-polarized scattering of p-polarized incident light.  This result is consistent with the absence of  satellite peaks in the result for $\left\langle \partial R_{\mathrm{pp}}/\partial\Omega_s \right\rangle_{\mathrm{incoh}}$ presented in Fig. 1(a).  The intensity maxima in the out-of-plane cross-polarized (p$\to$s) scattering intensity distribution depicted in Fig. 2(c) correspond to the peaks at $\theta_s \approx 19^{\circ}$ seen in the plot of $\left\langle \partial R_{\mathrm{sp}}/\partial \Omega_s \right\rangle_{\mathrm{incoh}}$ presented in Fig. 1(b).

The result that satellite peaks are observed in scattering processes in which the scattered light is s polarized, independent of the polarization of the incident light, is an interesting result of the present calculations.  It may be connected with the fact that s-polarized light is reflected more strongly from a dielectric surface than is p-polarized light.

In Fig. 3 we present results analogous to those presented in Fig. 1, but for angles of incidence $(\theta_0,\phi_0) = (5.19^{\circ}, 45^{\circ})$.  In Fig. 3(a) we present results for the in-plane $(\phi_s = \phi_0)$ co-(p$\to$p, s$\to$s) and cross-(p$\to$s, s$\to$p) polarized scattering, while in Fig. 3(b) we present results for out-of-plane $(\phi_s =\phi_0 \pm 90^{\circ})$ co-(p$\to$p, s$\to$s) and cross-(p$\to$s, s$\to$p) polarized scattering. In the results presented in Fig. 3(a) [3(b)] single-scattering processes give no contribution to the mean differential reflection coefficient for $-46.85^{\circ} < \theta_s < 65.57^{\circ}$ $(-54.59^{\circ} < \theta_s < 54.59^{\circ}$).  The limits of these angular regions are clearly seen in these figures.

A well-defined enhanced backscattering peak is seen in the results plotted in Figs. 3(a).  Satellite peaks are predicted by Eq.~(\ref{eq:13}) to occur (in-plane) at $\theta^{(1,2)}_{s} = -31.18^{\circ}$ and $19.68^{\circ}$ for p-polarized incident light, and at $\theta^{(1,2)}_s = - 23.20^{\circ}$ and $12.30^{\circ}$ for s-polarized incident light, when the angles of incidence were the same as in Fig.~3.  These scattering angles are indicated by vertical dotted lines in this figure.
Peaks at $\theta_s = -23.20^{\circ}$ and $\theta_s = 12.30^{\circ}$ are present in the s$\to$s co-polarized scattering contribution to the mean differential reflection coefficient.  There is no suggestion of peaks at $\theta_s = - 31.8^{\circ}$ and $19.68^{\circ}$ in the p$\to$p co-polarized scattering contribution to the mean differential reflection coefficient, nor any suggestions of peaks in the cross-polarized (p$\to$s, s$\to$p) contribution to it.  

\begin{figure}[p]
    \centering
    \includegraphics{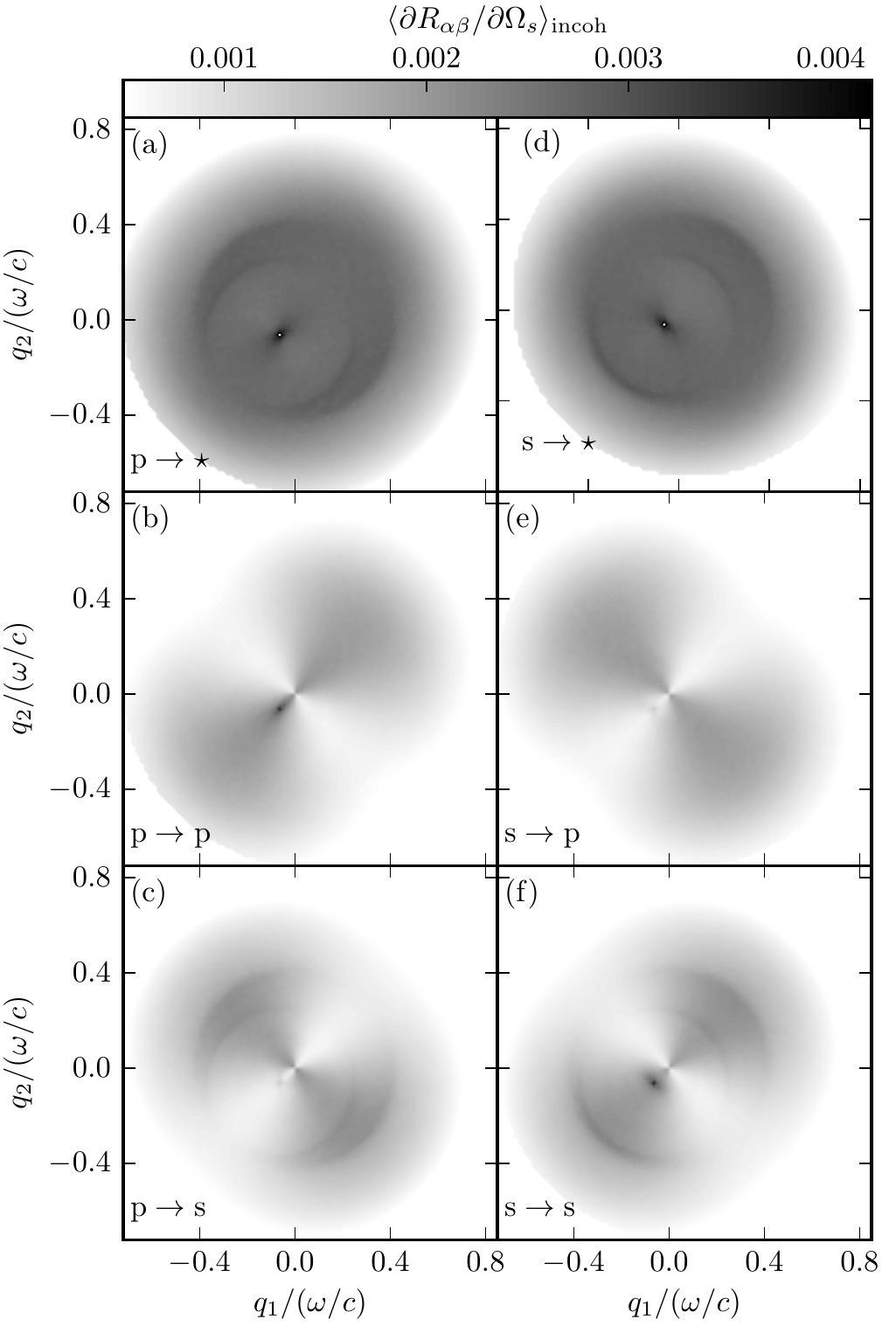}
    \caption{\label{fig:drc_2D_not_normal}
        The same as Fig. 2, but for angles of incidence given by $(\theta_0,\phi_0) = (5.19^{\circ}, 45^{\circ})$.
        Note that the color bar is cut at both ends in order to enhance the satellite rings. Also note that the contribution from single scattering is suppressed, i.e. the differential reflection coefficient is artificially set to $0$ for $\left|\pvec{q} - \pvec{k}\right| > k_-$.}
\end{figure}

The results for out-of-plane scattering presented in Fig. 3(b) show no enhanced backscattering peaks. The reason for this is simply that since the abscissa points along $\phi = \phi_0 \pm 90^\circ$, it does not cut through the backscattering peak, localized at $(\theta_s, \phi_s) = (\theta_0, \phi_0 + 180^\circ)$. We do see some remainders of the satellite ring structure; the low peaks around $\theta_s \approx \pm 20^\circ$ are part of the rings to the upper left in Fig.~\ref{fig:drc_2D_not_normal}. As the rings decay in strength away from the direction $\phi = \phi_0 \pm 90^\circ$, they are weaker than what is seen in-plane.

As a necessary, but not sufficient, condition of the validity of our simulation results is energy conservation. If all materials in the scattering system are lossless, i.e. $\mathrm{Im}(\varepsilon_i) = 0$ ($i=1,2,3$), the power of the scattered light has to be equal to the power of the incident light. Under these conditions, energy was conserved within 0.03\% in our simulations.

In conclusion, in this paper we have presented a nonperturbative approach to the solution of the reduced Rayleigh equation for the scattering of polarized light from a dielectric film with a two-dimensional randomly rough surface deposited on a planar metallic surface.  We have applied this result to calculate the contributions to the mean differential reflection coefficient  from the in-plane co- and cross-polarized components of the light scattered incoherently, as well as from the out-of-plane co-and cross-polarized components of the light scattered incoherently.  The out-of-plane scattering contributions have not been calculated in earlier perturbative studies of this problem \cite{8,11}.  In addition, we have calculated the full angular distribution of the intensity of the scattered light, which has helped to refine the conclusions drawn from the calculations of the mean differential reflection coefficient.  The main physical result obtained in this work is the demonstration that satellite peaks (or rings) can arise in scattering from the film structure studied here.  This result is in agreement with the results of Soubret \textit{et al.} \cite{8} but not with those of Kawanishi \textit{et al.} \cite{11}.  A detailed study of the conditions under which satellite peaks occur is lacking, but perhaps the approach developed here will be used to determine them.  The work reported here also opens the door to the possibility of calculating other properties of the light scattered from the film system studied here, such as all the elements of the associated Mueller matrix, and offers the possibility of designing such structures to possess specified scattering properties.  These are problems that have to be left to the future.

\section*{Acknowledgements}

The research of T.N., P.A.L., and I.S. was supported in part by NTNU by the allocation of computer time. The research of A.A.M. was supported in part by AFRL contract FA9453-08-C-0230.

T.N., P.A.L., and I.S. would like to thank Dr. Jamie Cole at the University of Edinburgh for his kind hospitality and fruitful discussions. They would also like to acknowledge the assistance of Dr. Fiona Reid and Dr. Christopher Johnson at the EPCC in parallelizing and optimizing the simulation code.

The work by T.N. and P.A.L. was partially carried out under the HPC-EUROPA2 project (project number: 228398) with the support of the European Commission---Capacities Area---Research Infrastructures.


\end{document}